\documentclass[aps,prl,twocolumn,showkeys,superscriptaddress,sectionbib,square]{revtex4-1}
\setcitestyle{numbers,square}
\usepackage{epsfig}
\usepackage{color}
\usepackage{times}
\usepackage{amsmath}
\usepackage{amssymb}
\usepackage{indentfirst}
\usepackage{graphicx}
\usepackage{bm}
\usepackage{wrapfig}
\usepackage{hyperref}
\usepackage{dcolumn}
\usepackage{epsfig}

\begin{document}
\title{Parameterization of the Stoner-Wohlfarth model of magnetic hysteresis}
\author{Nikolai~A. Zarkevich}\email{zarkev@ameslab.gov} 
\author{Cajetan Ikenna Nlebedim}\email{nlebedim@ameslab.gov}
\author{R.~William McCallum} 
\address{Ames Laboratory, U.S. Department of Energy, Ames, Iowa 50011-3020, USA}

\begin{abstract}
The Stoner-Wohlfarth is the most used model of magnetic hysteresis, but its computation is time-consuming.  
We use machine learning to approximate piecewise this model by easy-to-compute analytic functions.  
Our parametrization is suitable for fast quantitative evaluations and fitting experimental data, which we exemplify. 
\end{abstract}
\date{\today}
\maketitle

\section{\label{Introduction}Introduction}
\begin{figure}[t]
\begin{center}
\includegraphics[height=60mm]{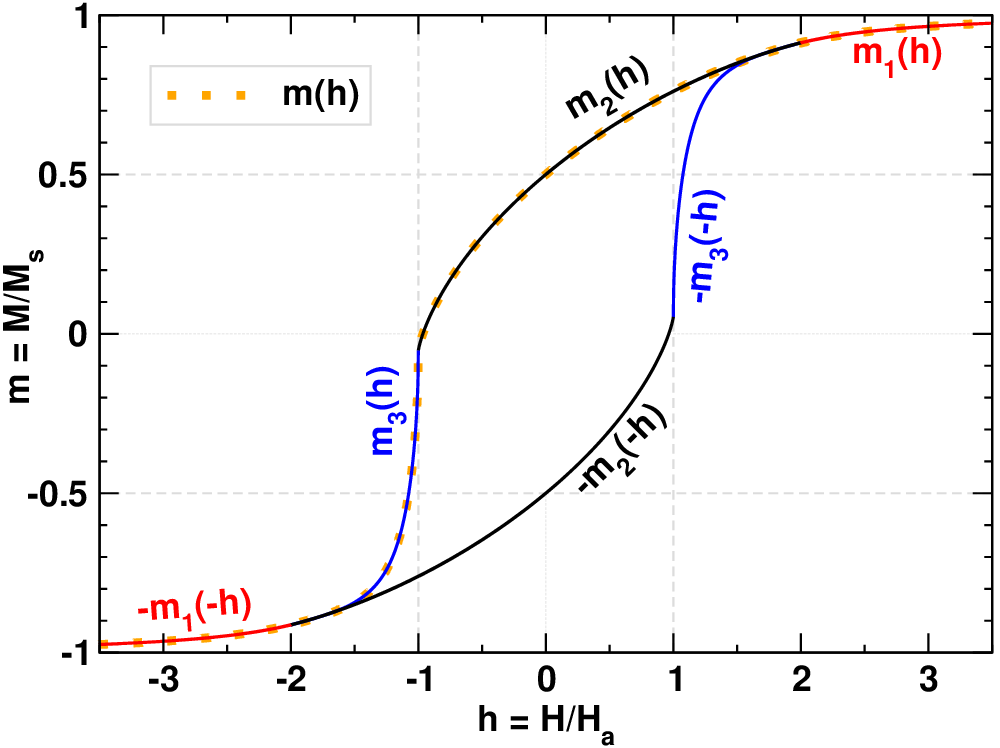}
\end{center}
\vspace{-2mm}
\caption{\label{FigSW} Piecewise parameterization of the SW hysteresis loop (orange dots) 
 by functions  
$m_1$, $m_2$, and $m_3$ (colored lines).  
}
\end{figure}

{\par }
Mathematical models \cite{StonerW1948,McCallum2005}, databases \cite{Complexity11p36y2006,AiiDA2016,OQMD2013}, and machine learning techniques \cite{CompMat4p25y2018,Nature533p73y2016,PRL92p255702y2004,PRB67p064104y2003,ActaMat50p2443y2002}
are extensively used for materials discovery \cite{JPhysD51n2p024002y2018,PRB93p020104y2016,SurfaceScience591n1pL292y2005,PRB71p115332y2005}.  
Known analytical approximations and predictive estimates \cite{PRL100p040602y2008,PRB75p104203y2007,PRB77p144208y2008,PRB89p134308y2014,JChemPhys142p024106y2015,JAC802p712y2019} greatly simplify those efforts, 
especially for magnetic materials \cite{APLMater2n3p032103y2014,PRB91p174104y2015,JChemPhys143p064707y2015,PRB97p014202y2018,Vonsovrkii1971}. 
There are several competing methods for approximating a function.  
\begin{enumerate}
\item
A smooth function can be approximated by a basis expansion. 
Examples are a Taylor expansion, a Fourier series, 
a basis of gaussians, etc. 
\item
Any piecewise-differentiable function can be  approximated piecewise by the rational functions, which may have poles.  An example is $1/r$.  
\item
Any continuous function can be approximated by a deep learning network (DLN), which is not differentiable \cite{Lapa1965,DeepLearning2016,AI3}.   
\end{enumerate}
{\par }
Preference is given to a more precise approximation with fewer fitted parameters. 
As a bonus, the rational functions and many analytic basis functions have known derivatives.  
Thus, for smooth or differentiable functions, methods 1 and 2 are preferable to DLN \#3. 
Combining a functional mapping with method 2, 
we approximate the Stoner-Wohlfarth (SW) piecewise-differentiable $m(h)$ curve (Fig.~\ref{FigSW}).  

{\par }
The SW  model \cite{StonerW1948} 
describes the hysteresis curve \cite{Stoletov1873}
for a random
distribution of non-interacting uniaxial particles whose magnetization
reverses through coherent rotation. 
It remains the most popular model of magnetic hysteresis for hard magnets \cite{McCallum2005}. 
The SW model \cite{StonerW1948}  presents
the magnetization curve $m(h)$ in terms of the reduced magnetization, $m=M/M_s$, where
$M$ is the magnetization and $M_s$ is the saturation magnetization at
infinite field, and $h=H/H_a$, where $H$ is the applied magnetic field and $H_a$ is the
anisotropy field of the material.  
Subsequent modifications \cite{McCallum2005} have
included the effect of interactions in the model. Unfortunately, the
SW model has no analytic solution, so the calculation of the SW
function requires the numerical integration of the $m$ vs. $h$ curves
for a distribution of particle orientations, where 
each individual $m_i (h)$ curve is obtained by minimizing the energy equations for
discrete values of $h$.  In real systems, the assumption of coherent
rotation invariably fails in the second quadrant, where either domain
wall motion or other modes of demagnetization (such as curling or
buckling) provide lower energy paths.  
Nevertheless, it has been demonstrated \cite{McCallum2005} 
that the $M(H)$ curves of hard magnetic
materials can be well described by a five-parameter fit with $M_s$, $H_a$,
an interaction parameter (demagnetization factor $\bar{\gamma}$), 
and the mean $H_S$ and width $\Delta H_S$ of a
switching field distribution (SFD).  
Furthermore, since the SW
assumptions are generally valid for the fitting of first quadrant
demagnetization curves, such fits yield accurate values of $M_s$ and $H_a$.
This is important in determining a detailed dependence of $M_s$
on temperature $T$ or composition $c$
 from experiment, 
since $M_s$ is often approximated by $M_{max}$ measured at
the highest  field $H_{max}$.  
Since $H_a (T)$ is a function of temperature $T$,
this results in an additional factor in $M(T)$.  The same is true when
the dependence of $M_s$ on composition is being
investigated in an alloy.

{\par}
While the calculation of the SW dependence $m(h)$ is straightforward, using the
tabulated values  to fit experimental data is cumbersome \cite{IEEE40p2907y2004,JPhaseED26n3p209y2005,JAP97n10p10M516y2005,PRB68p134452y2003}.  
The utility of the model can be greatly enhanced by an analytic parameterization
of the SW function $m(h)$, 
so that experimental data can be fitted easily.  
The SW data \cite{McCallum2005} was calculated 
to a precision of  7 significant digits in $m$ for steps of $0.001$ in $h \in [-20, 20]$, 
with known exact values of $m = \pm 1$ at $h = \pm \infty$ and $m(0)=0.5$, see Table~\ref{Tmodel}.  
The resulting numerical dependence $m(h)$ is then parametrized piecewise 
by three analytical functions 
with domains at 
$h \in [-2, -1)$, $[-1,2]$, and $[2,+\infty)$, 
see Fig.~\ref{FigSW}, 
and by two inverse analytical functions $h_{\pm} (m)$. 

{\par}
The SW model predicts a kink in the second derivative $d^2m/dh^2$ at $h= \pm 2$, which 
is well reproduced by the functions $m_1$ and $m_{(2)}^{\prime \prime}$, see Appendix. 

\section{Parametrization}
\subsection{Direct $m(h)$ }
{\par }
We approximate 
$m(h)$
by 3 analytic functions (Fig.~\ref{FigSW}):
\begin{equation}
\label{EqMHall} 
    m(h) = \left\{
 \begin{array}{c l } 
   m_1 (h),  &  h \in [2,+\infty); \\
   m_{2}  (h),          &  h \in [-1,2];   \\
   m_3 (h), &  h \in [-2, -1]. 
 \end{array}
  \right. 
\end{equation} 
	We denote $m_0 \! \equiv \! m(-1) \! \approx \! -0.05284686$, see Table~\ref{Tmodel}.
The whole hysteresis loop can be constructed using the inversion symmetry,
which transforms $m(h) \to -m(-h)$; for example, $m(h)=-m_1(-h)$ at $h \le -2$. 
At $h=-1$, the first derivative
$dm/dh$ is discontinuous and infinite  for $m_3$, 
but $(m_3(h)-m_0)^2$ has a finite slope for all $h \in [-2, -1]$, and we fit this squared function. 
At $h=2$, 
there is a well-known kink in the second derivative ${d^2m}/{dh^2}$,
 experimentally observed using the singular point detection techniques \cite{PhysicaB346p524y2004}.  

\begin{table}[b]
\caption{\label{Tmodel}
Values of $m(h)$ and $\overline{\cos \phi}$ from SW~\cite{StonerW1948}.  
}
\begin{tabular}{rll}
\hline
 $h$  & $m(h)$ & SW~\cite{StonerW1948} \\
\hline
$-\infty$ & $-$1 & $-$1 \\
$-$1  & $-$0.05284686 & $-$0.052631 \\
0 & $0.5$ & 0.5 \\
1 & 0.7607696 & 0.760770 \\
2 & 0.9129751 & 0.9130 \\
3 & 0.9645663 & 0.9646 \\
4 & 0.9809153 & 0.9809 \\
$\infty$ & 1 & 1 \\
\hline
\end{tabular}
\end{table}
{\par}
Using machine learning, we perform a piecewise least-squares (LS) fit of the following functions (with coefficients in Table~\ref{Tfit}):
\begin{eqnarray}
\left.\begin{aligned}
 \label{EqMh}
  m_1 (h \ge 2) &= \frac{1 + \sum_{n=1}^{3} c_n h^{-2n}  }{  1 + d_1 h^{-2} }  ,  \\
  m_2 (-1 \! \le \! h \! \le \! 2) \, &= \, \frac{0.5 + \sum_{n=1}^{6} c_n h^{n}}{1 + d_1 h + d_2 h^2}   ,\\  
  m_3(-2\! \le \!h\! \le \! -1) &= m_0 + \left[ \frac{\sum_{n=1}^{3} c_n (-h-1)^{n} }{ 1+ \sum_{k=1}^{3} d_k (-h-1)^{k} } \right]^{1/2}  
\end{aligned}\right.
\end{eqnarray}
These functions return 
$m_1(\infty) \equiv 1$, 
$m_2 (0) \equiv 0.5$, and
$m_3 (-1) \equiv m_0$, 
see Fig.~\ref{FigSW} and Table~\ref{Tmodel}.
Each function is accurate within its domain, see Appendix.  

\begin{table}[b]
\caption{\label{Tfit}
Coefficients of eqs.~\ref{EqMh} and accuracy of the fit: least-squares fit error $\chi^2$,  
Theil U, 
and correlation (C) coefficients.
}
\begin{tabular}{lrrrr}
\hline
           & $m_1$   & $m_2$ &  $m_3$ \\
\hline
$c_1$ & 18.2445 & 1.20029 & 2.90821 \\
$c_2$ & -5.96049 & 0.854124 & 10.1974 \\
$c_3$ & -3.1374 & 0.116601 & -5.12248 \\
$c_4$ & -- & -0.031271 & -- \\
$c_5$ & -- & 0.0045254 & -- \\
$c_6$ & -- & -0.00176557 & -- \\
\hline
$d_1$ & 18.5187 & 1.73361 & 4.01935 \\
$d_2$ & -- & 0.739904 & 13.3217 \\
$d_3$ & -- & -- & -7.55852  \\
\hline
$\chi^2$ & $5 \times 10^{-7}$ & $3 \times 10^{-6}$ & $2 \times 10^{-4}$   \\
U            & $5 \times 10^{-6}$ &  $4 \times 10^{-5}$ & $1 \times 10^{-3}$  \\
C        & 1.000000      & 1.000000         & 0.999995             \\
\hline
\end{tabular}
\end{table}
\begin{figure}[t]
\begin{center}
\includegraphics[height=60mm]{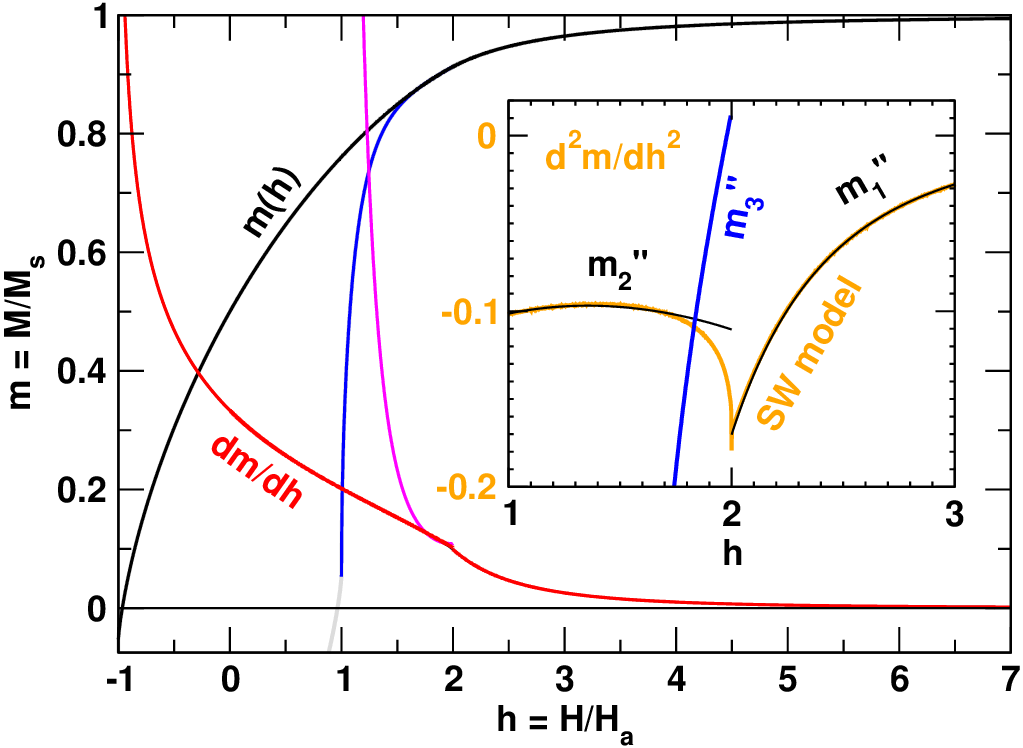}
\end{center}
\vspace{-2mm}
\caption{\label{Fig1} The SW model (increasing grey $m(h)$ lines), 
its first (decreasing orange $dm/dh$ in the plot), and second derivatives (orange $d^2m/dh^2$ lines in the inset),  
and our approximation by the analytic functions: 
black $m_1$ and $m_2$, blue $-m_3$, as well as 
their first (decreasing, same scale, red $m_1'$ and $m_2'$, pink $-m_3'$ lines) and (inset) second derivatives (black $m_1''$ and $m_2''$, blue $-m_3''$ lines). 
Except for the second derivatives (inset), the SW model and the analytic functionals are indistinguishable, 
thus most of the gray and orange lines (SW model) are covered by the approximating lines.  
}
\end{figure}
\subsection{Inverse $h(m)$ }
{\par}
We also parametrize the inverse function by
\begin{equation}
 \label{EqHM}
    h(m) = \left\{
 \begin{array}{c l }
   h_+ (m > m_0),  &  h \in [-1,+\infty); \\
   h_- (m \le  m_0), &  h \in (-\infty, -1]. 
 \end{array}
  \right.
\end{equation} 
A single analytical function $h_+ (m)$ in the first quadrant 
ignores a kink in $d^2m/dh^2$ at $h=2$, but covers both $m_2$ and $m_1$. 
Similarly, $h_- (m)$ is the inverse function for both $-m_1 (-h)$ and $m_3 (h)$, see Fig.~\ref{FigSW}. 
{\par}
First, we map $h \in (-\infty, +\infty)$ onto $y \in [0,1]$ 
using the transformation $y = 1/[(h+1)^2 +1]$, and piecewise fit $y(m)$ by two rational functions: 
one increasing and one decreasing. 
Next, we substitute those into the inverse transformation 
$
  h = -1 + sqrt( abs( -1 + 1/y ))
$, 
where $sqrt(x)=\sqrt{x}$ is the square root and 
$abs(x)=|x|$ is the absolute value (needed only near $h=-1$). 
We get
\begin{eqnarray}
 \label{EqHMp}
  h_{\pm} (m) = -1 + \left| -1+\frac{1+\sum_{n=1}^{5} {c_n (m \mp 1)^n} }{\sum_{k=1}^{5} {d_k (m \mp 1)^k}}   \right|^{1/2}   
\end{eqnarray}
with coefficients in Table~\ref{TableHM}. 
Here $h_+$ is expanded in terms of $(m-1)$. 
The error in $y[h(m)]$ with $h_{\pm}(m)$ approximated by eq.~\ref{EqHMp} is below $6 \times 10^{-4}$.
\begin{table}[b]
\caption{\label{TableHM}
Coefficients of eqs.~\ref{EqHM}, see Table~\ref{Tfit} caption.
}
\begin{tabular}{lrrrr}
\hline
           & $h_+$   & $h_-$ \\
\hline
$c_1$ &   -16.0382            &  -9.14903  \\
$c_2$ &    33.7616            & 61.4734  \\
$c_3$ &    106.516            &  -336.135  \\
$c_4$ &     91.6034           &  0.290317  \\
$c_5$ &          --       &  18813.8  \\
\hline
$d_1$ &    -2.59778            & 4.55815     \\
$d_2$ &     0.782093           & 0.577252    \\
$d_3$ &     -43.4887           &  -357.282    \\
$d_4$ &     -52.8431            &   209.511    \\
$d_5$ &     -41.8217            &  18676.5     \\
\hline
$\chi^2$ & $7 \times 10^{-5}$   & 0.0001  \\
U           &  0.0004   & 0.0005  \\
Corr.     & 1.000000  & 1.000000  \\
\hline
\end{tabular}
\end{table}
{\par}

\begin{figure}[t]
\begin{center}
\includegraphics[width=75mm]{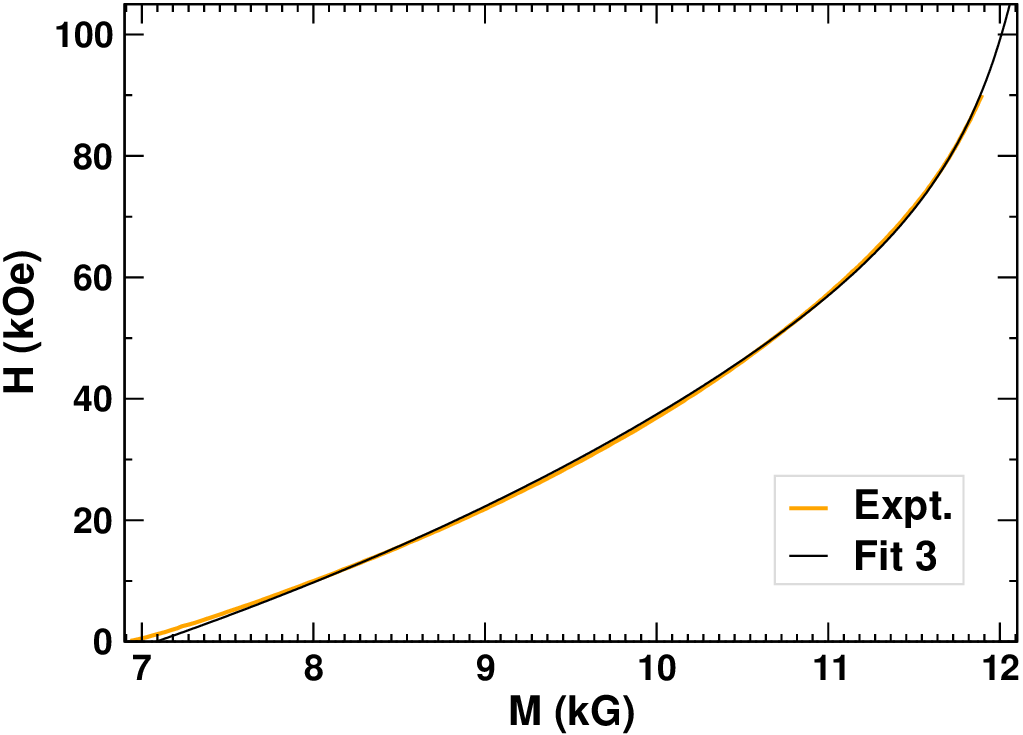}  
\end{center}
\vspace{-2mm}
\caption{\label{FigExpt} Experimental data (Expt., orange), 
fitted (Fit 3, black line) by 3-parameter eq.~\ref{EqHdemag} at $H \! > \! 0$.  
}
\end{figure}
\begin{figure}[t]
\begin{center}
\includegraphics[width=80mm]{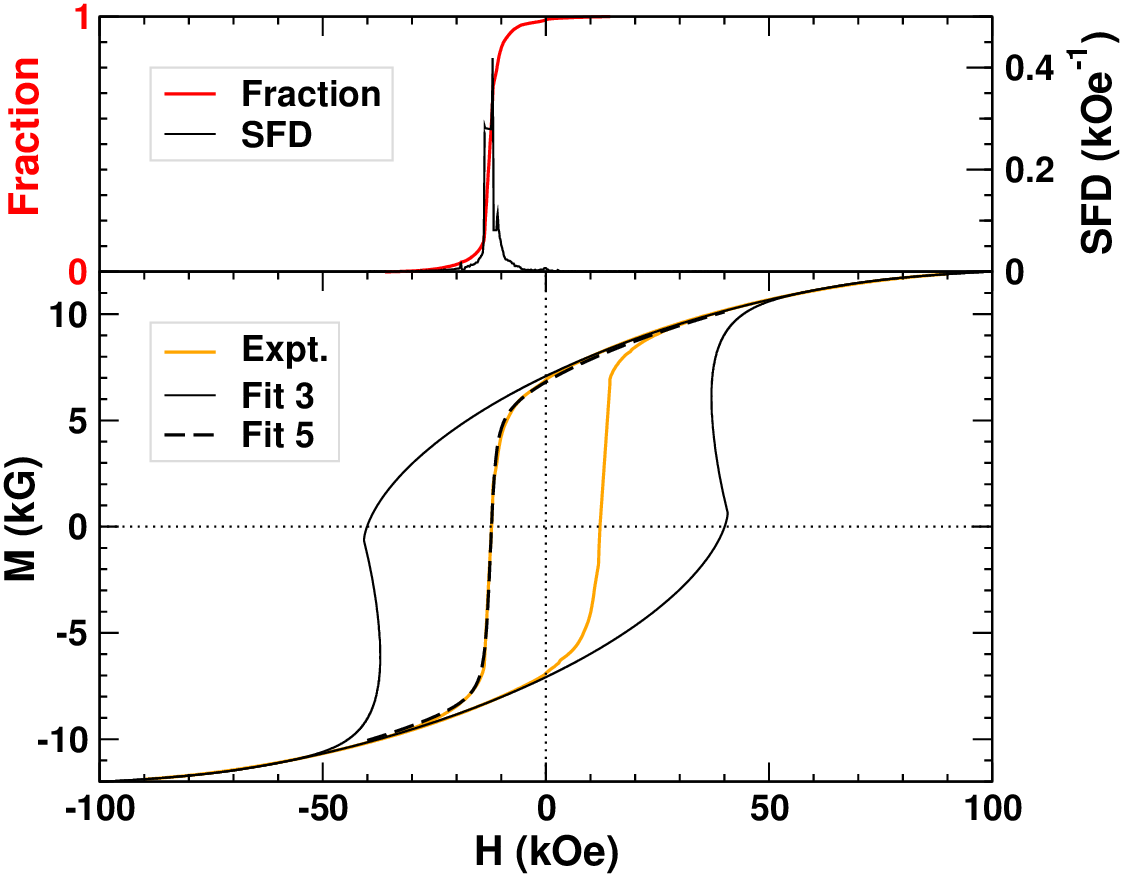}       
\end{center}
\vspace{-2mm}
\caption{\label{FigSFD} (Lower) Experimental M(H) data (Expt., orange lines), 
its 3-parameter fit (Fit 3,  solid black lines) by the 
theoretical curves with demagnetization (eq.~\ref{EqHdemag}), 
and 5-parameter fit (Fit 5,  dashed black line) with SFD approximated by eq.~\ref{eqSLdH}. 
(Upper) A fraction of the reversed magnetization (left scale, red) and its derivative (SFD: right scale, black).  
}
\end{figure}

\section{Application}

{\par}
{\bf Experiment}. 
The magnetic hysteresis loop was measured for the ribbons of an Ames rare-earth magnetic alloy. 
To prepare the ribbons, ingots with composition (Nd$_{0.80}$Pr$_{0.20}$)$_{2}$Fe$_{14}$B were prepared by arc melting materials of constituent elements in argon atmosphere. Melt spun ribbons were prepared by inductively melting the ingots in quartz crucibles and ejecting the melt onto a single copper wheel at 30 m/s surface velocity through a 0.8 mm orifice. Melt spinning was performed in 1/3 atmosphere of high purity He gas. The as-spun ribbons were crystallized by heat treatment at 700$^{\circ}$C for 15 min in 1/3 ultra-high purity argon atmosphere. Magnetic hysteresis loop was measured at 300 K in a Quantum Design vibrating sample magnetometer with maximum applied magnetic field of 90 kOe. 
\\

{\par}
{\bf Analysis}. 
Eq.~\ref{EqHMp} can be used to fit a measured data at $H\!>\!0$, taking into account demagnetization:  
\begin{equation}
 \label{EqHdemag}
  H (M) = H_a h_+ (M/M_s) - \bar{\gamma} M ,
\end{equation}
where $\bar{\gamma} = \gamma / 4 \pi$ is a demagnetizing factor.
Using a sufficiently large initial guess of $M_s$ to avoid a singularity of $H(M)$ at $M=M_s$, 
we fit experimental  $H(M)$ data 
at $H \! > \! 0$ by 3 parameters ($H_a$, $M_s$, $\bar{\gamma}$) in eq.~\ref{EqHdemag}.

{\par}
The result of fitting the measurements at $H>0$ by eq.~\ref{EqHdemag} is in Fig.~\ref{FigExpt}.
We find  
$H_a=41.55\,$kOe, $M_s = 12.54\,$kG, and $\bar{\gamma} = 1.2$. 

{\par}
	Due to the switching field, demagnetization data deviates from the SW model (or its parametrization) at $H<0$. 
A fraction of the reversed magnetization  can be approximated by a sigmoid curve $S(x)$
with two parameters ($H_S$ and $\Delta H_S$) in the argument $x = (H-H_S)/\Delta H_S$.  

{\par}
We use $S_L (x)=\frac{1}{2}(L(x)+1)$ with 
the classical Langevin function
$L(x) = coth(x) - 1/x$.
Its first derivative 
is a bell-shaped curve:  
\begin{eqnarray}
\label{eqSLdH}
 & S_L^{\prime} (H)  \equiv \frac {d S_L}{d H} = \frac{1}{2} \frac {dL}{dx} \frac {dx}{dH} =  \hfill  \\
 & = \frac{1}{2 \, \Delta H_S^L} \left[ 1 - coth^2 \left( \frac{H-H_S}{\Delta H_S^L} \right) + \left( \frac{\Delta H_S^L}{H-H_S} \right)^2 \right]  \nonumber
\end{eqnarray}

{\par }
	Comparing the upper branch of the experimental $M(H)$ data to the two branches [$m(h)$ and $-m(-h)$] 
of the parametrized SW model, 
we get a fraction of the reversed magnetization,  see Fig.~\ref{FigSFD}. 
By fitting it to $S_L$, 
we get 
 $H_S = -12.44\,$kOe and $\Delta H_S^L = 0.484 \,$kOe. 
Its derivative is the switching field distribution (SFD), 
which can be approximated by 
eq.~\ref{eqSLdH}.   


\section{\label{Summary}Summary}
{\par} 
We have provided a convenient analytic approximation for the Stoner-Wohlfarth  model \cite{StonerW1948}, 
suitable for quick and easily computations. 
We applied it to the measured magnetic hysteresis loop 
and fitted the experimental data    
 by 5 parameters: 
$M_s$, $H_a$,  $\bar{\gamma}$,   $H_S$, and $\Delta H_S$. 
Our easy-to-compute analytic functions  serve as useful tools for 
description of magnetic materials, that facilitate materials discovery \cite{JPhysD51n2p024002y2018,ActaMat154p365y2018,ActaMat173p225y2019,ActaMat180p341y2019}. 


\section*{Acknowledgments}
{\par} 
We thank Professors Ivan~I. Oleynik and Duane~D.~Johnson for inspiration and advising. 
This work was supported by the U.S. Department of Energy, Office of Basic Energy Sciences, Division of Materials Science and Engineering. The research was performed at the Ames Laboratory, which is operated for the U.S. DOE by Iowa State University under contract DE-AC02-07CH11358.  

\appendix 
\section{\label{aprecision} 
Analysis of $m(h)$ fit and its derivatives}

{\par}
The fitted function $m(h)$ and its derivatives are shown in Fig.~\ref{Fig1}. 
Difference between $m(h)$ and its fit by the analytic functions $m_n$ ($n=1,2,3$) from eq.~\ref{EqMh} is within the numeric noise.
The first derivative of $m_n$ 
is indistinguishable from $dm/dh$ for the model. 
The inset in Fig.~\ref{Fig1} shows that 
$d^2m/dh^2$ is reproduced well at $h > 2$ or $h<1.7$, 
but not at $h =2-\delta$. 
The 2nd derivative of $m_1$ (at $h \ge 2$) coincides with the model, 
while $d^2 m_2 / dh^2$ deviates at $h =2-\delta$ for $0 \le \delta < 0.3$, 
and the 2nd derivative of $-m_3(-h)$ (blue line in Fig.~\ref{Fig1} inset) has a larger deviation near $h=2$.

{\par}
The function $m(h)$ itself and its first derivative are reproduced very well everywhere, 
hence the lines for the SW model and its analytic approximation are indistinguishable in Fig.~\ref{Fig1}. 
The function $m_1 (h)$ and its derivatives are as good as the tabulated values of $m(h)$ at $h\ge 2$. 
The largest error (calculated as a deviation from $m(h)$ at a given $h$) is 
$3 \times 10^{-5}$ for $m_1$ at $h=2$ 
(it is $<\! 1 \times 10^{-5}$ for $h>2.537$ and  
   $<1.34 \times 10^{-5}$ for $h>2.066$);
0.001 for $m_3$ at h=0 ($<0.0005$ at $0.12<h<1$);
and 
0.0008 for $m_{2}$ at $h=-1$ (0.0001 at h=2, and $\le \! 1 \! \times \! 10^{-4}$ at $-0.986 <h \le 2$). 

{\par}
The first derivative at $h=2$ is 
$m^{\prime} (2)  \equiv dm/dh |_{h=2} =0.09869$ for the model. 
The error in $m^{\prime} (2)$ constitutes 
0.00037 for $m_1^{\prime}$,
0.003  for $m_2^{\prime}$, and  
0.01 for $m_3^{\prime}$.  
This error in $dm/dh$ is the largest for $m_1^{\prime}$ and $m_3^{\prime}$, 
while 
$dm_2/dh$ has a smaller error $\le 0.001$ at $-0.8<h<1.9$, 
but an expectedly large error exceeding 0.01 near $h=-1$. 

{\par}
The second derivative $m_1^{\prime \prime} \equiv d^2 m_1/dh^2$ reproduces the model correctly 
in the whole domain of $m_1$ at $h \in [2, +\infty)$, see Fig.~\ref{Fig1} inset. 
However,  $m_2^{\prime \prime}$ and $m_3^{\prime \prime}$ 
 deviate from the model at $h = 2-\delta$, 
where  $0 \le \delta <0.3$. 
At $h=2$,  this deviation reaches
0.006 for $m_1^{\prime \prime}$,
0.06  for $m_2^{\prime \prime}$, and
0.19 for $m_3^{\prime \prime}$ second derivative, 
where $m^{\prime \prime} (2) \equiv d^2m/dh^2 |_{h=2} \approx -0.18$ for the model, see Fig.~\ref{Fig1} inset. 
If needed, the second derivative $d^2 m/dh^2$  at $-1<h\le 2$ can be directly approximated by eq.~(7): 
\begin{eqnarray}
m_{(2)}^{\prime \prime} = 
-0.303321 \, (h+1)^{-1} 
-0.156365
-0.049342 \, h \nonumber \\
+0.0893185 \, h^2-0.0458726 \, h^3+0.0118055 \, h^4   \nonumber \\
 + 0.148387 \, (2-h)^{1/2} + 0.0524192 \, (2-h)^{1/4}  ,
\nonumber
\label{Eq2deriv}
\end{eqnarray}
which is accurate even at $h=2-\delta$.
The function $m_{(2)} \equiv \int \int m_{(2)}^{\prime \prime}$, obtained by 
double integration of $m_{(2)}^{\prime \prime}(h)$ at $-1<h\le 2$, 
 contains roots and a logarithm, which are slower to compute. 

{\par} 
Equations \ref{EqMh} and \ref{Eq2deriv} provide accurate analytic expressions for $m(h)$ function and its 1st and 2nd derivatives (a luxury, not offered by DLN), 
while the inverse function $h(m)$ in eq.~\ref{EqHMp} facilitates a fit of experimental data.

\bibliography{SW}
\end{document}